\documentclass[conference]{IEEEtran}

\usepackage{amsmath}
\usepackage{amssymb}
\usepackage{booktabs}
\usepackage{graphicx}
\usepackage{url}
\usepackage{cite}

\begin{document}
%
\title{Optimal Resource Allocation with Node and Link Capacity Constraints in Complex Networks}

\author{\IEEEauthorblockN{Rui Li, Yongxiang Xia}
\IEEEauthorblockA{College of Information Science and Electronic Engineering\\ Zhejiang University, Hangzhou, 310027, China\\
Email: xiayx@zju.edu.cn}
\and
\IEEEauthorblockN{Chi K. Tse}
\IEEEauthorblockA{Department of Electronic and Information Engineering\\
Hong Kong Polytechnic University, Hong Kong, China\\
Email: encktse@polyu.edu.hk}
}

\maketitle

\begin{abstract}
With the tremendous increase of the Internet traffic, achieving the best performance with limited resources is becoming an extremely urgent problem. In order to address this concern, in this paper, we build an optimization problem which aims to maximize the total utility of traffic flows with the capacity constraint of nodes and links in the network. Based on Duality Theory, we propose an iterative algorithm which adjusts the rates of traffic flows and capacity of nodes and links simultaneously to maximize the total utility. Simulation results show that our algorithm performs better than the NUP algorithm on BA and ER network models, which has shown to get the best performance so far. Since our research combines the topology information with capacity constraint, it may give some insights for resource allocation in real communication networks.
\end{abstract}


%
\IEEEpeerreviewmaketitle

\section{Introduction}
With the rapid development of the communication technology, the number of Internet users increases sharply. Meanwhile, the Internet traffic has exploded recently in response to new multicast and multimedia applications. These all require a larger Internet capacity and how to achieve the best transportation performance with limited resources is becoming an overwhelming problem.

Although we can increase the network capacity by enlarging the delivery capacity of nodes or bandwidth of links, it is quite expensive to do so \cite{xiang2013traffic}. Instead, the recent development of complex network theory provides a new way to investigate this problem. Recently, Du et al. firstly explored the intersection between complex networks and optimization theory, and used the improved optimization theory to solve real-world problems in networked systems\cite{du2015adequate,du2016identifying}. This emerging topic soon draws general attention from diverse communities. In our case, it is possible to optimize the resources allocation in complex communication networks to improve the whole network capacity.

In real communication networks, communication resources mainly include bandwidth resource \cite{xiang2013traffic}, node processing capabilities \cite{liu2015optimal,xia2008internet} and buffer resource \cite{xue2013effect}. Researchers tried to optimize the allocations of these resources to enhance the communication performance. However, some significant constraints in real communication networks have not been taken into account. For example, the research in \cite{liu2015optimal} only considers the capacity constraint of nodes without considering the capacity constraint of links. Actually, link capacity constraint is related to the link bandwidth, which is an important constraint in real communication network. Therefore, it should be considered in the optimization problem as well. Moreover, the flow rate should be also optimized to enhance the communication performance \cite{low1999optimization}. Based on these concerns, to achieve better performance, this paper will consider optimal capacity resource allocation and flow control simultaneously.

\section{Resource Allocation Algorithm}
\subsection{Traffic Model}
Consider a network that consists of $N$ nodes, $L$ links and $S$ traffic flows flowing from source nodes to destination nodes. $V=\{1,2,\textellipsis,N\}$ denotes the set of nodes, $E=\{1,2,\textellipsis,L\}$ denotes the set of links, and $F=\{1,2,\textellipsis,S\}$ denotes the set of flows. For flow $s$ , $V(s)$ represents the set of nodes flow $s$ goes through, $E(s)$ represents the set of links flow $s$ passes. Obviously, $V(s)\subseteq V$ and $E(s)\subseteq E$.

Assume node $n$ has capacity $D_{n}$. We assume the total capacity of the whole network is fixed, i.e., $\Sigma_{n}D_n=M$. Meanwhile, we assume link $l$ has capacity $C_l$, and traffic flow $s$ has rate $x_s$. $x_s$ is restricted in the interval $I_s=[m_s,M_s]$ \cite{low1999optimization}, given $m_s\geqslant0$ and $M_s < \infty$.

$F(l)=\{s\in F|l\in E(s)\}$ refers to the set of traffic flows passing link $l$. Similarly, $F(n)=\{s\in F|n\in V(s)\}$ refers to the set of traffic flows going though node $n$.  When the aggregate flow rate going through node $n$ is no more than the capacity of node $n$, and the aggregate flow rate passing link $l$ is no more than the capacity of link $l$, the network is in the stable state; otherwise, the network is in congestion state. We use $U_s$ to denote the utility of flow $s$. Our work aims to find the best capacity resource allocation method of nodes and links to achieve the largest utility under the premise of avoiding congestion.

\subsection{Optimization Problem}
Under the condition that the total capacity of nodes is fixed to $M$, to avoid congestion, resource allocation problem is formulated as:
\begin{equation}
\max_{x_s\in I_s, C_l\geqslant0, D_n\geqslant0}    \sum \limits_s U_s(x_s)  \qquad  \qquad  \qquad \qquad
\end{equation}
\begin{equation}
 subject\ to \qquad \sum \limits_{s\in F(n)} x_s \leqslant D_n, \ n=1,2,\ldots,N
\end{equation}
\begin{equation}
\qquad \qquad \qquad \sum \limits_{s\in F(l)} x_s \leqslant C_l, \ l=1,2,\ldots,L
\end{equation}
\begin{equation}
\sum \limits_{n} D_n = M
\end{equation}
Here, $U_s$ represents the utility function. In many congestion-management mechanisms such as TCP Reno and TCP Vegas, $U_s$ is increasing and strictly concave. Assume link $l_{mn}$ connects nodes $m$ and $n$. The larger the total capacity of $m$ and $n$ is, the higher the possibility that link $l_{mn}$ has a large capacity exists. We're inclined to assume there is a positive correlation between the total capacity that node $m$ and $n$ possess and the capacity that link $l_{mn}$ possesses. So we can establish the following equality relation \cite{xia2008internet}
\begin{equation}
C_{l_{mn}}=T_{mn}\alpha(D_m+D_n)
\end{equation}
where $T$ is the adjacency matrix of the network, in which $T_{mn}=T_{nm}=1$ if nodes $m$ and $n$ are connected, and $T_{mn}=T_{nm}=0$ otherwise. $\alpha$ is the proportional constant. Notice that $T_{mn}$ contains the topology information of the network. So with equation (5), we combine the topology information of the network with the optimal resource allocation problem, and inequation (3) can be rewritten as:
\begin{equation}
\sum \limits_{s\in F(l_{mn})} x_s \leqslant T_{mn}\alpha(D_m+D_n), \ m,n=1,2,\ldots,N
\end{equation}

The above problem can be viewed as a convex optimization problem, and the primal optimal solution must exist \cite{boyd2004convex}. However, it is almost impossible to solve the primal problem directly in real networks, since it requires coordination among possibly all sources \cite{low1999optimization}. So, we first look at its dual. Define the Lagrangian
\begin{equation}
\begin{aligned} 
L(x,D;p,q,\lambda) = \sum \limits_s U_s(x_s) - \sum \limits_{n=1}^N q_n(\sum \limits_{s\in F(n)}x_s - D_n) \qquad \qquad \qquad \qquad \qquad \\ - \sum \limits_{m=1}^N \sum \limits_{n=1}^N p_{mn}(\sum \limits_{s\in F(l_{mn})}x_s - T_{mn}\alpha(D_m+D_n))\qquad \qquad \qquad \qquad  \\ - \lambda(\sum \limits_{n=1}^N D_n -M) \qquad \qquad \qquad \qquad \qquad \qquad \qquad \qquad \qquad \qquad \ \  \\
\end{aligned}
\end{equation}
Here, we introduce variables $p_{mn}$, $q_n$ and $\lambda$. We denote $p_{mn}$ as the price link $l_{mn}$ pays to transmit one unit traffic flow, and $q_n$ as the price node $n$ pays per unit traffic flow. Since the network we considered here is undirected, $l_{mn}$ and $l_{nm}$ refers to the same link connecting nodes $m$ and $n$, so the price $p_{mn} = p_{nm}$. Transforming the above equation, we get
\begin{equation}
\begin{aligned} 
L(x,D;p,q,\lambda) = \sum \limits_{m=1}^N \sum \limits_{n=1}^N p_{mn}T_{mn}\alpha(D_m+D_n) + \sum\limits_{n=1}^N q_nD_n  \qquad \\ +  \sum \limits_s [ U_s(x_s)-x_s(\sum \limits_{m=1}^N \sum_{\substack{n=1 \\ l_{mn}\in E(s)}}^N p_{mn}+\sum \limits_{n\in N(s)}q_n) ] \qquad \\  - \lambda(\sum\limits_{n=1}^ND_n - M) \qquad \qquad \qquad \qquad \qquad \qquad \qquad \qquad
\end{aligned}
\end{equation}
When it reaches the primal optimal solution, the flow rate $x_s$ and the node capacity $D_n$ satisfy
\begin{equation}
\frac{\partial L(x,D;p,q,\lambda)}{\partial x_s} = 0
\end{equation}
\begin{equation}
\frac{\partial L(x,D;p,q,\lambda)}{\partial D_n} = 0
\end{equation}
According to the above two equations, we get
\begin{equation}
U_s^{'}(x_s) = \sum \limits_{m=1}^N \sum_{\substack{n=1 \\ l_{mn}\in E(s)}}^N p_{mn} + \sum \limits_{n\in V(s)}q_n
\end{equation}
\begin{equation}
\lambda = q_n + 2\sum\limits_{m=1}^N p_{mn} T_{mn} \alpha
\end{equation}
With $p_{mn}=p_{nm}$, and we can further transform (12) into
\begin{equation}
\lambda = \frac{1}{N}\sum\limits_{n=1}^N (q_n + 2\sum\limits_{m=1}^N p_{mn} T_{mn} \alpha)
\end{equation}
The objective function of the dual problem is thus
\begin{equation}
D(p,q,\lambda) = \max_{x_s\in I_s, D_n \geqslant0} L(x,D;p,q,\lambda)
\end{equation}
In order to maximize $L(x,D;p,q,\lambda)$, variables $x_s$ and $\lambda$ must satisfy (11)-(13). Therefore, after we eliminate intermediate variable $\lambda$, we have
\begin{equation}
\begin{aligned}
D(p,q) = \sum \limits_{s}[U_s(x_s)-x_s(\sum\limits_{m=1}^N \sum_{\substack{n=1 \\ l_{mn}\in E(s)}}^N p_{mn} + \sum\limits_{n\in N(s)}q_n)] \\
+ \frac{1}{N}\sum\limits_{n=1}^N q_nM + \frac{2}{N}\sum\limits_{m=1}^N \sum\limits_{n=1}^N p_{mn}T_{mn}\alpha M \qquad \ \ \
\end{aligned}
\end{equation}
and the dual problem becomes
\begin{equation}
\min_{p\geqslant0,q\geqslant0} D(p,q)
\end{equation}
According to (11), we have
\begin{equation}
x_s = [U_s^{'-1}(\sum\limits_{m=1}^N \sum\limits_{n=1}^N p_{mn} + \sum\limits_{n\in V(s)}q_n)]_{m_s}^{M_s}
\end{equation}
where $[z]_a^b = min\{max\{z,a\},b\}$, and $U_s^{'-1}$ is the inverse of $U_s^{'}$. We use the gradient projection method \cite{low1999optimization} to solve the dual problem
\begin{equation}
p_{mn}(t+1)=p_{mn}(t)-\gamma(\frac{\partial D}{\partial p_{mn}}(p_{mn}(t),q_{n}(t)))
\end{equation}
\begin{equation}
q_{n}(t+1)=q_n(t)-\gamma(\frac{\partial D}{\partial q_{n}}(p_{mn}(t),q_n(t)))
\end{equation}
Here, $\gamma$ refers to the stepsize. Then, we compute $\frac{\partial D}{\partial p_{mn}}$ and $\frac{\partial D}{\partial q_{n}}$. Notice that $x_s$ is the function of $p_{mn}$ and $q_n$. According to the chain rule for derivatives, we have
\begin{equation}
\frac{\partial D}{\partial p_{mn}} = -\sum\limits_{s\in F(l_{mn})} x_s + \frac{2}{N}\sum\limits_{m=1}^N \sum\limits_{n=1}^N \alpha T_{mn} M
\end{equation}
\begin{equation}
\frac{\partial D}{\partial q_n} = -\sum\limits_{s\in F(n)} x_s + \frac{1}{N} M
\end{equation}
So, we have
\begin{equation}
p_{mn}(t+1)=p_{mn}(t)+\gamma(\sum \limits_{s\in F(l_{mn})} x_s -\frac{2}{N}\sum\limits_{m=1}^N \sum\limits_{n=1}^N\alpha T_{mn}M)
\end{equation}
\begin{equation}
q_{n}(t+1)=q_n(t)+\gamma(\sum \limits_{s\in F(n)} x_s -\frac{1}{N}M)
\end{equation}
With equations (22) and (23), we assign different price for different nodes and links.

\subsection{Distributed Algorithm}
We design the following iterative algorithm based on the above derivation.
\begin{itemize}
\item Initial State
\subsubsection{Link price for $l_{mn}$}
$p_{mn}(0)\geqslant0$;
\subsubsection{Node price for $n$}
$q_{n}(0)\geqslant0$;
\subsubsection{Flow rate for $s$}
$x_s(0)\in I_s$;
\end{itemize}
\begin{itemize}
\item Link $l_{mn}$'s Algorithm \\
At time $t$, link $l_{mn}$ receives flow rate $x_s(t)$, $s\in F(l_{mn})$, and adjusts its transmission price according to
\begin{equation}
\nonumber
\begin{aligned}
p_{mn}(t+1)=[p_{mn}(t)+\gamma(\sum\limits_{s\in F(l_{mn})}x_s(t) \\ - \frac{2}{N}\sum\limits_{m=1}^N \sum\limits_{n=1}^N \alpha T_{mn}M)]^{+} \qquad
\end{aligned}
\end{equation}
\end{itemize}
\begin{itemize}
\item Node $n$'s Algorithm \\
At time $t$, node $n$ receives flow rate $x_s(t)$, $s\in F(n)$, and adjusts its price according to
\begin{equation}
\nonumber
q_n(t+1)=[q_n(t)+\gamma(\sum\limits_{s\in F(n)}x_s(t) - \frac{1}{N}M)]^{+} \qquad
\end{equation}
Meanwhile, node $n$ adjusts its capacity according to
\begin{equation}
\nonumber
D_n(t)=\sum\limits_{s\in F(n)}x_s(t)
\end{equation}
\end{itemize}
\begin{itemize}
\item Flow $s$'s Algorithm \\
At time $t$, flow $s$ receives the link price $p_{mn}(t)$, $l_{mn}\in E(s)$, and node price $q_n(t)$, $n\in V(s)$, and adjusts its transmitting rate according to
\begin{equation}
\nonumber
x_s(t+1)=[U_s^{'-1}(\sum\limits_{m=1}^N \sum_{\substack{n=1 \\ l_{mn}\in E(s)}}^N p_{mn}(t) + \sum\limits_{n\in V(s)}q_n )]_{m_s}^{M_s}
\end{equation}
\end{itemize}

\section{Simulations}
\subsection{Network Models}
To verify our algorithm, we consider two theoretical network models, including Barab¨¢si-Albert (BA) scale-free network model \cite{barabasi1999emergence} and Erdos-Renyi (ER) random network \cite{erdos1961evolution} both with $N=200$ nodes and $L=600$ links. We guarantee that each and every network is connected. In each network, $S=1000$ pairs of nodes are randomly selected as sources and destinations. Once a pair of nodes is chosen, we assume there is a flow from the source to the destination along the shortest path. If there are more than one shortest path, then we randomly choose one.

Firstly, we need to test the convergence of our distributed algorithm. Compare the result of our algorithm with optimal value solved by MATLAB CVX package \cite{grant2008cvx}, with
\begin{equation}
error=\sqrt{\frac{\sum\limits_{s}(x_s^{\ast}-x_s^{cvx})^2+\sum\limits_{n}(D_n^{\ast}-D_n^{cvx})^2}{S+N}}
\end{equation}
Here, $(x_s^{\ast}, D_n^{\ast})$ is the final iteration value of our algorithm, and $(x_s^{cvx},D_n^{cvx})$ is the solution of MATLAB CVX package.

The simulation results in Table 1 are averaged over 10 realizations with $M=5000$, $\alpha=0.1$. As we can see, the computational error is so small that can be ignored.

\begin{table}
\caption{Error}
  \centering
  \begin{tabular}{lll}
    \hline
    BA & ER \\
    \noalign{\global\arrayrulewidth1pt}\hline\noalign{\global\arrayrulewidth0.4pt}
    3.6921e-06 & 7.2048e-07\\
    \hline
  \end{tabular}
\end{table}
\subsection{Performance Indicators}
\begin{figure}
\centering
\includegraphics[width=0.43\textwidth]{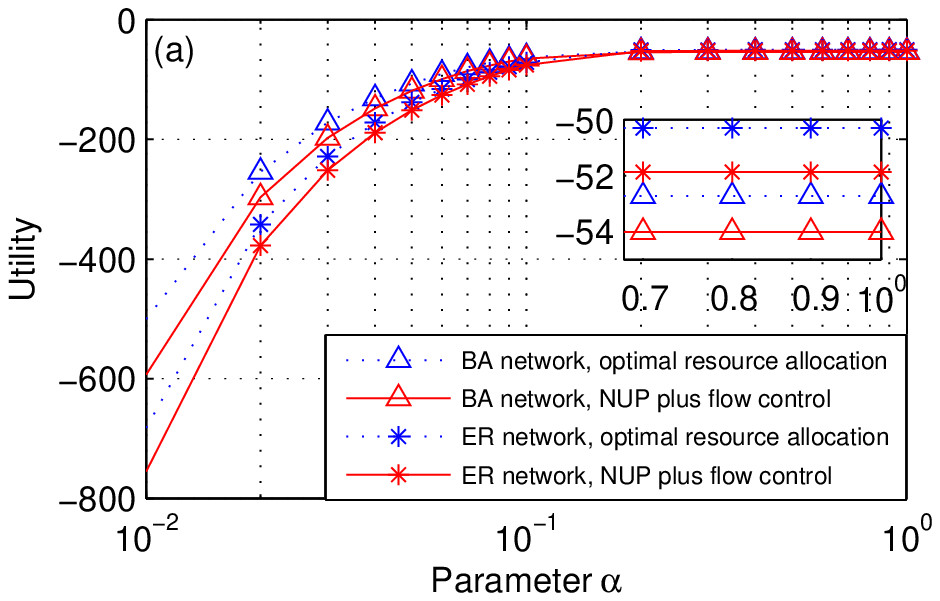}
\includegraphics[width=0.43\textwidth]{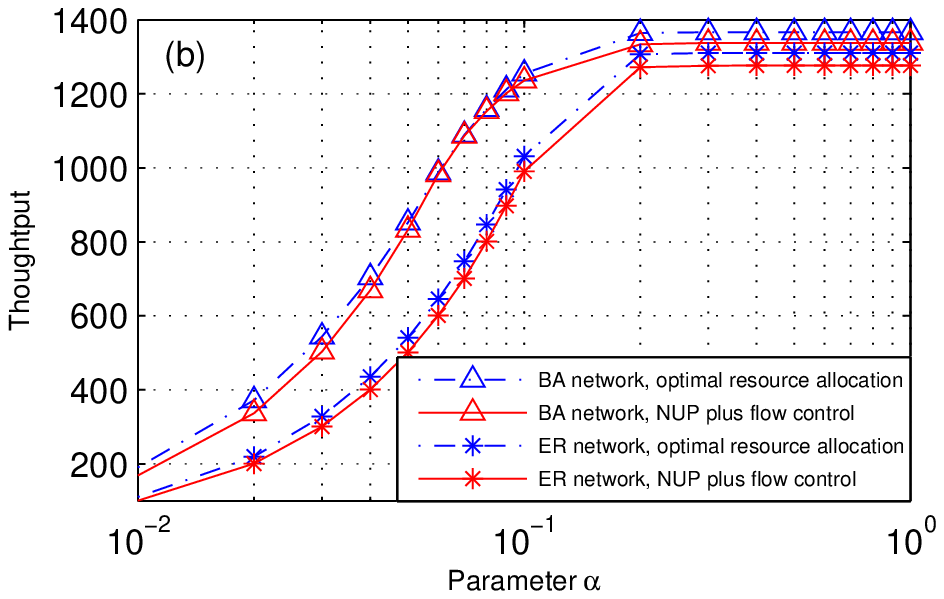}
\caption{Performance of BA and ER networks as a function of parameter $\alpha$. $M=5000$. All the results are averaged over 10 simulations.}
\label{fig.1}
\end{figure}

\begin{figure}
\centering
\includegraphics[width=0.43\textwidth]{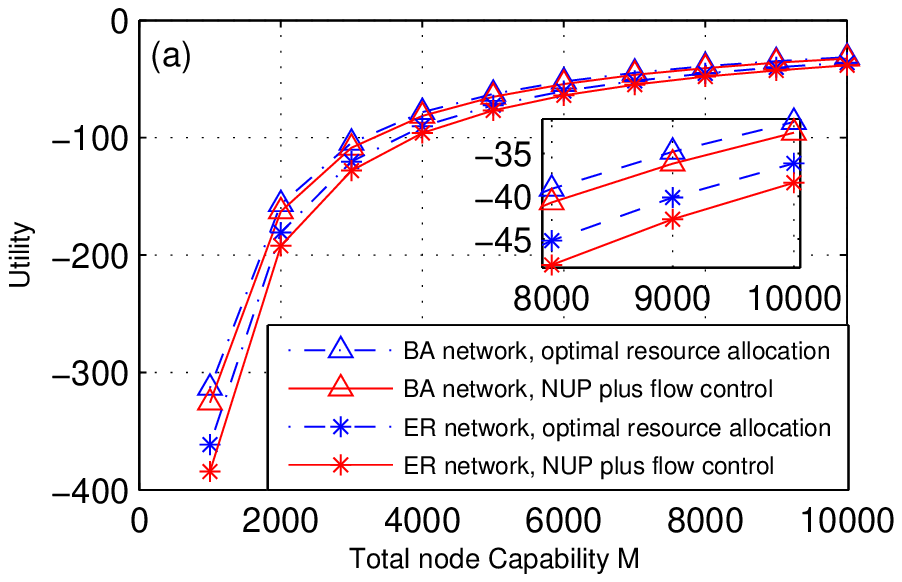}
\includegraphics[width=0.43\textwidth]{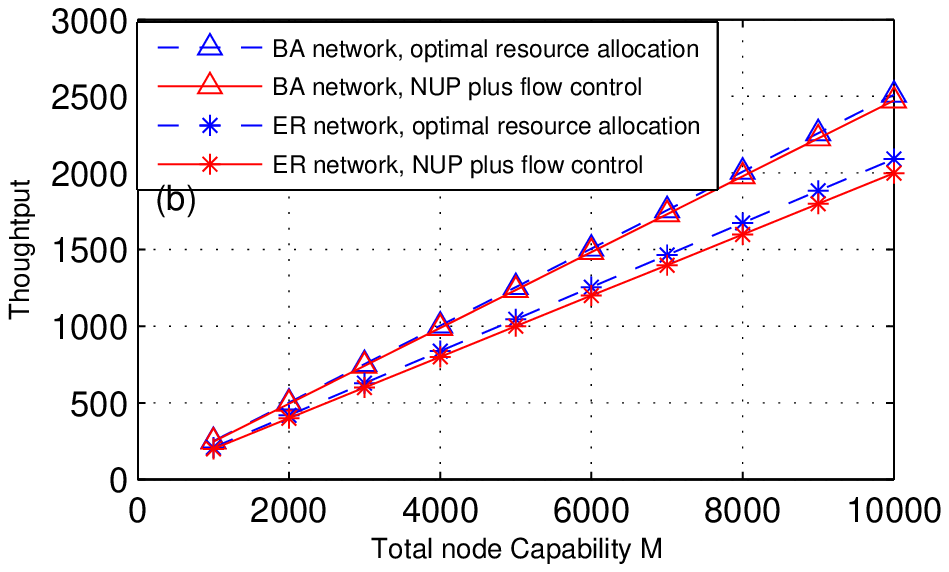}
\caption{Performance of BA and ER networks as a function of total node capacity M. $\alpha=0.1$. All the results are averaged over 10 simulations.}
\label{fig.2}
\end{figure}
Here, we measure the network performance by the following two aspects: total utility and network throughput. Firstly, because the original purpose is to find the maximum utility, we choose $\sum\limits_{s}U_s(x_s)$ as our first performance indicator. We use $U_s(x_s)=\frac{-1}{d_s^2x_s}$ as the utility function, where $d_s$ refers to the number of nodes flow $s$ passes. This definition is a simplified version used in TCP Reno. The other indicator we employ is throughput, given $throughput=\sum\limits_{s}x_s$. The larger throughput the network has, the better performance the network shows.
\subsection{Simulation Results}
The solution of our algorithm is compared with recent resource allocation algorithm named NUP\cite{wu2013analysis}, which is defined as
\begin{equation}
U(n)=\frac{\sum_{\substack{u,w\in V \\ u\neq w\neq n}}\sigma_{uw}(n)}{\sum_{n\in V}\sum_{\substack{u,w\in V \\ u\neq w\neq n}}\sigma_{uw}(n)}
\end{equation}
\begin{equation}
D_n=MU(n)
\end{equation}
where $\sigma_{uw}(n)$ is the total number of paths from node $u$ to $w$ meanwhile pass through node $n$. However, NUP algorithm is only a resource allocation algorithm, and we can't get the flow rate $x_s$. So, we combine NUP with an optimal flow control algorithm proposed by Low and Lapsley \cite{low1999optimization}, and compare the result of our algorithm with that combination.

Figures 1 and 2 show that our algorithm has larger utility and throughput compared with NUP algorithm. In Fig.1, we can see clearly that when total node capacity $M$ is fixed to $5000$, the utility and throughput are nondecreasing functions of parameter $\alpha$. As $\alpha$ increases, the link capacity constraint is loosed up, and the whole system gains better performance. However, merely increasing $\alpha$ when it's large enough may have little benefit because total node capacity $M$ will become a bottleneck, so we can see in Fig.1, performance almost remains unchanged after $\alpha=0.2$.

In Fig. 2 , network utility and throughput keep increasing along with the growth of total node capacity $M$. Obviously, it is in line with our cognition. As the total node capacity $M$ increases, with equality relation (5), the capacities of links and nodes increase simultaneously. Thus we can increase flow rate without congestion. As a result, the utility and throughput both increase with the increasement of flow rate $x_s$.

Since our algorithm allocates the capacities of nodes and links with the target to maximize the utility of flows, while NUP only allocates the capacity of nodes in proportional to the node usage probability, the performance of our algorithm is better than NUP. Moreover, NUP can only do a part of job as our algorithm can do. It has to be accompanied with an optimal flow control algorithm to maximize the utility of flows. In this way, our algorithm shows its advantage.

\section{Conclusion}
In this paper, we design a iterative algorithm based on Duality Theory to solve the optimal resource allocation problem. Our algorithm considers both the constraints of node and link capacity resources, which makes the algorithm fit the realistic condition better. Simulation results show that our algorithm performs better than the existing best resource allocation algorithm named NUP. Our work may give some advice to resource allocation in real communication networks.




%
\bibliographystyle{IEEEtran}
\bibliography{IEEEabrv,IEEEexample}

\end{document}